\documentclass[aps,prl,twocolumn]{revtex4}

\usepackage{graphicx}

\newcommand{\beq}{\begin{equation}}
\newcommand{\enq}{\end{equation}}
\newcommand{\bea}{\begin{eqnarray}}
\newcommand{\ena}{\end{eqnarray}}
\newcommand{\bd}{b^{\dagger}}
\newcommand{\rr}{{\mathbf r}}
\newcommand{\tc}{t_{\mathrm c}}

\newcommand{\refeq}[1]{Eq.~(\ref{#1})}
\newcommand{\reffig}[1]{Fig.~\ref{#1}}

\begin{document}
\title{Transition from a two-dimensional superfluid to a 
one-dimensional Mott insulator}
\author{Sara Bergkvist}
\affiliation{Department of Theoretical Physics, Royal Institute of Technology,
AlbaNova, SE-106 91 Stockholm, Sweden}
\author{Robert Saers}
\affiliation{Department of Physics, Ume{\aa} University, 
SE-90187 Ume{\aa}, Sweden}
\author{Emil Lundh}
\affiliation{Department of Physics, Ume{\aa} University, 
SE-90187 Ume{\aa}, Sweden}
\author{Anders Rosengren}
\affiliation{Department of Theoretical Physics, Royal Institute of Technology,
AlbaNova, SE-106 91 Stockholm, Sweden}
\author{Magnus Rehn}
\affiliation{Department of Physics, Ume{\aa} University, 
SE-90187 Ume{\aa}, Sweden}
\author{Anders Kastberg}
\affiliation{Department of Physics, Ume{\aa} University, 
SE-90187 Ume{\aa}, Sweden}

\begin{abstract}
A two-dimensional system of atoms in an anisotropic optical lattice is
studied theoretically.  If the system is finite in one direction, it
is shown to exhibit a transition between a two-dimensional superfluid
and a one-dimensional Mott insulating chain of superfluid tubes.
Monte Carlo (MC) simulations are consistent with the expectation that the
phase transition is of Kosterlitz-Thouless (KT) type.  The effect of
the transition on experimental time-of-flight images is discussed.
\end{abstract}


\maketitle

Cold atomic gases in optical lattices provide a means to study
many-body quantum phenomena that offers both versatility, precision,
and control.  These techniques allowed for the spectacular realization
of 
the Mott transition by Greiner {\it et al}.\ \cite{greiner2002},
which has inspired numerous proposals for creating and detecting
exotic quantum phases in such systems.

By manipulating the alignment and the polarizations of the laser beams
that build up the optical lattice, the geometry of the lattice can be
changed. The probability for tunneling can be adjusted by changing the
separation between potential wells, or by simply varying the
irradiance. In conjunction, this brings about the possibility of
designing optical lattices where the probability for tunneling differs
significantly between different directions, and where this difference
in tunneling rates can be a control parameter. By allowing the atoms
to tunnel only along one direction, one essentially has created a 2D
lattice of 1D quantum gases. By allowing tunneling in two directions,
an array of 2D systems is created~\cite{kohl2005}.

The tunability allows to explore the crossover between different
dimensionalities. Thus, with a properly chosen geometry, it should be
possible to start off with high laser irradiance and thus with deep
potential wells. This system will be a Mott insulator,
where the number of particles per site is fixed to an integer and
there is no phase coherence.  By lowering the irradiance 
in one direction, 
increasing
the tunneling probability, one should reach a point where tunneling
becomes probable along just one direction. This would give us isolated
1D tubes, known as Luttinger liquids
\cite{giamarchi2004}. Decreasing the irradiance 
in a second direction,
we expect to reach a point where there is a crossover from this to a
system of 2D superfluids, and eventually one global 3D
superfluid~\cite{ho2004,cazalilla2006,gangardt2006}. In this Letter,
this dimensional crossover is addressed by quantum MC simulations of
an anisotropic bosonic Hubbard model in two dimensions. We show how
the model can be readily realized in experiment, and
predict the outcome of absorption imaging on both sides of the
transition.

{\it System.} -- Consider a gas of bosonic atoms at zero 
temperature in a 2D, anisotropic optical lattice. 
The gas is described by the Bose-Hubbard Hamiltonian \cite{jaksch1998} 
\bea
H &=&-\sum_{i}\left( t_x\bd_{i_x,i_y} b_{i_x+1,i_y} +t_y\bd_{i_x,i_y} b_{i_x,i_y+1}+h.c.\right) \nonumber\\
&& +\frac{U}2 \sum_{i} n_i(n_i-1)-\mu\sum_i n_i,
\label{hamiltonian}
\ena 
where $i=(i_x,i_y)$;  $b_i$, $\bd_i$ are bosonic annihilation and
creation operators, $n_i=\bd_i b_i$,
$U$ the interaction
strength, $\mu$ the chemical potential, and $t_x$, $t_y$ the
tunneling matrix elements in the Cartesian directions.  The
crossover from 2D to 1D behavior considered in this paper can be
realized in a simple cubic optical lattice with a potential of the
form $ V(\rr) = \sum_{\alpha=x,y,z} V_{\alpha} \cos^2 k \alpha$. 
If the potential barrier in the $z$ direction $V_z$ is strong enough, 
there can be no tunneling in the $z$ direction, and
the sample can be considered as a stack of independent 2D systems
\cite{kohl2005}. 
$V_x$ should be chosen relatively weak in order to allow for superfluid 
flow in the $x$ direction, and $V_y$ is scanned over the 
physically interesting range. 
As will be shown, if the lattice has 
an extent of eight sites in the $x$ direction 
there will occur a phase transition 
for $t_x/U=0.3$ and $t_y/U\approx 6\times 10^{-3}$. As an example, for
a gas of $^{87}$Rb atoms in an optical lattice with wavelength 
$\lambda=850$ nm, these tunneling matrix elements are obtained by
choosing potential strengths $V_x=2.5 E_{\rm rec}$, $V_y=30E_{\rm
rec}$, and $V_z=80E_{\rm rec}$, where 
$E_{\rm rec}$ is the
single-photon recoil energy for a wavelength of 780nm \cite{greiner2002}.

{\it Phase diagram.} -- The Hubbard model exhibits a zero-temperature
quantum phase transition from 
a superfluid to a Mott insulating state 
when the ratio of the tunneling matrix elements and the coupling
strength, $t_x/U$ and $t_y/U$, are decreased below a critical value
\cite{fisher1989}. 
In one dimension (i.e., either $t_x=0$
or $t_y=0$), it is known that the phase transition occurs at $t/U=0.3$
for the system with an average of one particle per site
\cite{kuhner2000}.  For the anisotropic 2D model considered here, a
mean-field argument implies that $t$ should be replaced by the sum
$t_x+t_y$ in the 1D expression, so the transition occurs when
$(t_x+t_y)/U=0.3$. For finite systems the phase transition occurs at a
smaller value of $t/U$. 

\begin{figure}
\includegraphics[width=0.94\columnwidth]{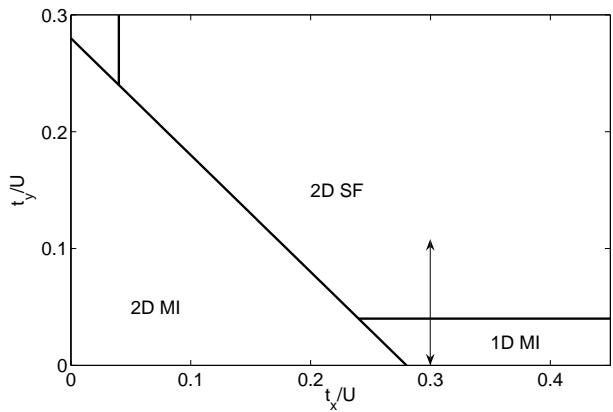}
\caption[]{Phase diagram for the finite anisotropic Hubbard model at number 
density $n=1$. 
The 2D superfluid and Mott insulating phases are denoted 
by ``2D SF'' and ``2D MI'', respectively. 
The size of the 1D Mott insulating phase ``1D MI'' 
has been exaggerated in the figure. 
The vertical double arrow indicates the interval along which 
the MC calculations in this study have been performed.
\label{fig:phasediagram}}
\end{figure}
Consider the anisotropic case, where one of the tunneling matrix
elements, say, $t_x$, is larger than the critical 1D value and the
other one, $t_y$, is varied.  In an infinite 2D system, it is known
that the system remains a 2D superfluid (2D SF) right down to the
point where $t_y$ vanishes \cite{efetov1974}. In other words, when the
coupling in one direction vanishes, the system can be seen as an array
of uncoupled Luttinger liquids.  However, any finite tunneling, no
matter how small, will induce phase coherence between the Luttinger
liquid tubes, so that an anisotropic phase,
with superflow only in one direction, does not exist.  However, the
situation changes when the system is not infinite. If each Luttinger
liquid has a finite length, there is an energy barrier against
tunneling into and out of it, and as a result there exists, for small
$t_y$, an anisotropic state with coherence only along the $x$
direction \cite{ho2004,cazalilla2006,gangardt2006}. Hence, when the
weaker tunneling matrix element $t_y$ is small enough, the state of
the system is a 1D Mott insulator (1D MI), where each site is a Luttinger
liquid tube extending along the $x$ direction.  
If the tunneling $t_y$ is
increased above a critical value $t_c$, the tubes become phase
correlated and the system undergoes a quantum phase transition to a
superfluid state. In a geometry where the system is infinite in the
weakly coupled direction, the state is an 1D SF, 
i.~e., a superfluid chain of Luttinger liquids, 
just above the phase
transition, but is expected to undergo a crossover into a 2D SF as the
coherence length becomes smaller than the tube length $L_x$, and the
picture of separate Luttinger liquid tubes can no longer be
sustained. The crossover between 1D SF and 2D SF presumably happens
over a very short interval, why we in the following shall speak of the
1D MI - 2D SF transition. In an actual experiment where the system is
finite, there is a crossover between 1D MI and 2D SF without an
intermediate 1D SF state.  In the 1D MI phase, each tube along the $x$
direction behaves as an independent Luttinger liquid, characterized by
an (inverse) exponent $K$ \cite{giamarchi2004,ho2004,cazalilla2006}.
This so-called Luttinger parameter determines the algebraic decay of
the single-particle correlation function $\Gamma(j)$,
according to
\beq
\label{correlation}
\Gamma_x(j) \equiv \langle \bd_{i_x,i_y} b_{i_x+j,i_y}\rangle \sim
j^{-1/(2K_x)}.  \enq This form for the asymptotic behavior holds only
in the 1D case, i.~e., for decoupled Luttinger liquid tubes.  
$\Gamma_y$ is defined analogously.
The critical tunneling was in Refs.\
\cite{ho2004,cazalilla2006,gangardt2006} found to depend on the
parameters as
\beq
\label{tc}
\tc \propto L_x^{-2+\frac1{2K_x}}. 
\enq 
Since it is known that $K_x > 2$ here \cite{kuhner2000}, the
exponent in \refeq{tc} is expected to be slightly less than -1.75.
The constant of proportionality in Eq.~(\ref{tc}) depends on the model used to integrate
out the dynamics in the tubes as well as on the dimensionality. These
phase boundaries have been indicated at arbitrarily chosen positions
in the phase diagram in Fig.~\ref{fig:phasediagram}.

\begin{figure}
\includegraphics[width=0.96\columnwidth]{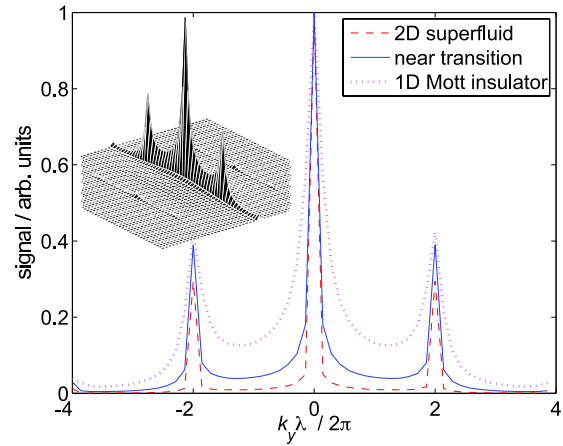}
\caption{Velocity distribution along the weakly coupled lattice
direction.  The curves corresponds to a ratio of the tunneling matrix
element to the interaction strength, $t_y/U=0.012$ (dashed line),
0.0025 (full), and $1.5\times 10^{-3}$ (dotted). The inset shows the
full 2D profile of the velocity distribution for the case
$t_y/U=1.5\times 10^{-3}$. 
The system size is 12$\times$12 sites.}
\label{fig:tof}
\end{figure}

{\it Experimental signal.} -- Detection of the transition can be achieved by applying an optical lattice potential, $V(\rr)$, to a Bose-Einstein condensate in a magnetic trap.
Ramping up $V(\rr)$ adiabatically ensures that the atoms are not heated, and 
that they are 
loaded to the lowest energy band of the lattice \cite{greiner2002}. 
For high lattice barriers, the correlations between wells are lost after some
hold time, resulting in a Mott insulating state, whereas in the case of low barrier heights the sample stays coherent.
The state of the system is detected by standard time of flight absorption imaging. When the
sample is released by ramping down the lattice and magnetic trap non-adiabatically, it will expand freely. 
The expansion reveals the sample's velocity distribution
$n(\mathbf{k}$), given by the Fourier transform of the trapped sample
density distribution in real space $n(\mathbf{r})$
\cite{gerbier2005prl}, $n(\mathbf{k})\propto \left\vert w(\mathbf{k})\right\vert^2S(\mathbf{k})$,
where the first part is the Fourier transform of the Wannier
function for the first Bloch band $w(\rr)$, and the second part, the
structure factor $S(\mathbf{k})$, is the Fourier transform of the
correlation function 
$\Gamma(\rr)$ defined 
in \refeq{correlation}.
The structure factor has the periodicity $4\pi/\lambda$.

Knowing the structure factor from MC simulations (see below), and
calculating the Wannier functions for a cubic lattice, a theoretical
prediction for the actual experimentally observable time-of-flight
profiles can thus be obtained.  The result is shown in
\reffig{fig:tof}.  The profile of the cloud in the direction of weak 
tunneling is determined by the state of the system. When distinct 
peaks are visible at $k=0, \pm4\pi/\lambda$, a 2D~SF is detected. 
This is connected with a slow decay of the real-space correlation 
function $\Gamma(\rr)$.
If the
lattice depth is increased, the sample passes into the
1D~MI and the interference fringes are blurred 
as $\Gamma(\rr)$ becomes exponentially decaying. 
The signal would be even more pronounced for a larger system.
However, the peaks
will be visible far into the Mott insulator regime, an effect caused
by maintained short-range coherence~\cite{gerbier2005prl}.

{\it{Method. -- }} To solve for the ground state of the Bose-Hubbard
model, \refeq{hamiltonian}, Quantum MC calculations with the
stochastic series expansion algorithm are used \cite{SaPRB99,SaPRE02}.
We simulate a system of size $L_x \times L_y$ lattice sites with
periodic boundary conditions to minimize edge effects.  
The inverse temperature is $\beta=(k_BT)^{-1}=1000U^{-1}$; 
this is large enough to ensure that the system is in its ground state. 
The results presented in the article
are calculated with a fixed $t_x=0.3U$ with $t_y/U$ being varied
across the anticipated phase transition as indicated by the double
arrow in \reffig{fig:phasediagram}. 
Trial calculations not shown
here, using larger values of $t_x/U$, have yielded similar results.
The calculations have been performed at the fixed average number of
particles per site $n=1$.

{\it{Transition. -- }} 
To locate the 2D~SF$\leftrightarrow$1D~MI
transition, three different observable quantities are calculated, each
of which can be considered a measure of superfluidity. 
The superfluid density, $\rho_{s}$, is readily obtained from the 
simulation data as $\rho_{s}=\langle W_y^2\rangle/\beta$, where 
$W_y^2$ is the so-called square winding number in the $y$ direction, 
defined as the squared net number of times a particle 
crosses the periodic boundary in the calculations \cite{SaPRB04}.
Second, the off-diagonal correlation function along the $y$ direction, 
$\Gamma_y(j)$, is, as we have seen, 
experimentally accessible through the momentum distribution, 
which is closely related to its Fourier transform. We now study the
correlation function in real space evaluated at half the system size,
$\Gamma_y(y=L_y/2)$ which we expect to behave differently above and
below the transition. Above, $\Gamma_y(y)$ decays algebraically, below
exponentially.  A third observable is the compressibility,
defined as the mean of variances,
$\Delta N^2 = \sum_{i_y} \mathrm{var} \left(\sum_{i_x}n_{i_x,i_y}\right)/L_y$.
It is expected that $\Delta N^2$ changes its behavior at the
transition: due to the finite length of the tubes, the fluctuations in
the number of particles per tube will be suppressed once the system
becomes insulating in the weakly coupled $y$ direction.

\begin{figure}
\begin{center}
  \resizebox{70mm}{!}{\includegraphics{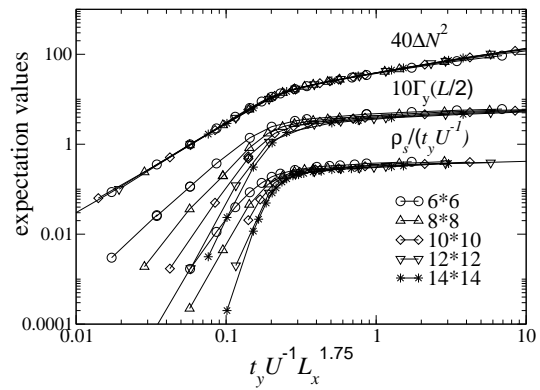}}
\end{center}
\caption{From top to bottom, the density fluctuation, 
the correlation, and the superfluid density are shown as a function of
the size-independent parameter $t_yL_x^{1.75}/U$, for different system 
sizes $L_x\times L_y$ measured in lattice sites, 
as indicated in the legend. The scale is logarithmic 
on both axes.
The correlation function and
the compressibility are multiplied by arbitrarily chosen 
numerical factors to separate the
three groups of lines.
The superfluid density is divided by the tunneling matrix element 
for better visual clarity.
}
\label{fig:Three}
\end{figure}

In \reffig{fig:Three}, the three quantities, $\rho_{s}$, $\Gamma_y(L_y/2)$,
and $\Delta N^2$, are shown for lattices $6\le L_x=L_y\le 14$.
The observables are
plotted as functions of $t_yL_x^{1.75}/U$. The exponent of $L_x$ is
given by \refeq{tc} and the value of $K_x$ is determined by a
finite-size scaling of the superfluid density as explained below.  
For a system with $t_x/U=0.4$, not shown here, the best scaling is
obtained for $t_yL_x^{1.79}/U$. 
The superfluid density in the $y$
direction is in the superfluid phase 
seen to be approximately 
proportional to the coupling 
constant $t_y/U$, hence the superfluid density is divided by $t_y/U$.
It is clear from \reffig{fig:Three} that there is a change in behavior
around $t_y/U \approx 0.3L_x^{-1.75}$.
The three quantities in \reffig{fig:Three} are behaving as in a 1D MI-SF 
transition. The true 1D system can be obtained by putting $L_x=1$ and letting 
$L_y\to\infty$, and we then recover the known value for the true 1D 
transition $t_y/U=0.3$ \cite{kuhner2000}. 
(The same curves are obtained for $L_y > L_x$, but these data are not 
shown here not to overload the figure). The finite size effects above the 
transition point are taken care of by scaling $L_x$ with the power 1.75. 
If the curves instead were plotted as a function of $t_y/U$, they 
would not tend to coincide.

Since in the critical region, the system can be described as a 1D
chain of Luttinger liquid tubes, the phase transition is a 1D quantum
phase transition.  Such a transition can be mapped onto a classical
phase transition in 2D \cite{Girvin}, and if the number of particles
is held
fixed as in the present study, it is known that the transition
considered here is a KT transition
\cite{fisher1989}. 

\begin{figure}
\includegraphics[width=0.85\columnwidth]{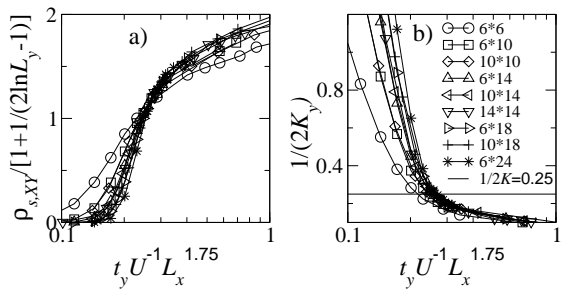} 
\caption{a) Superfluid density in the dual XY model, 
\refeq{eq:rhoscale}, 
scaled according to Weber-Minnhagen scaling. 
b) Exponent 
of the algebraic decay of the correlation $\Gamma_y$, 
as a function of the dimensionless
coupling parameter $t_yL_x^{1.75}/U$. The horizontal line shows the
critical value expected for a 1D Mott transition at fixed density,
$1/(2K)=0.25$. Different curves correspond to different system sizes;
the same symbols are used in both parts of the figure.}
\label{fig:scale}
\end{figure}

Characteristic for such a transition is a discontinuous jump of the
superfluid density at the critical point \cite{SaPRB04}.  In
\reffig{fig:scale} a) a finite size scaling of the data for the
superfluid density is illustrated.  In order to do the scaling, a
series of mappings has to be performed: First, according to Ref.\
\cite{ho2004}, the original Hubbard model can be rewritten as a
Josephson junction chain model, where each Luttinger liquid tube is
treated as a site, with a charging energy $E_C\propto L_x^{-1}$ and
hopping energy $E_J\propto t_yL_x^{1-1/2K_x}$.  Next, this model is in
turn mapped onto a 2D XY-model if one identifies
$\beta_{XY}=\sqrt{E_J/E_C}$, $L_{x,XY}=\beta\sqrt{E_JE_C}$, and
$L_{y,XY}=L_{y}$ \cite{Girvin}, where the quantities referring to the
dual 2D XY-model carry the subscript XY. As a result, the superfluid
density in the 2D XY-model is \cite{SaPRB04}
\begin{equation}
\rho_{s,XY}\propto \langle W_y^2\rangle L_{y}/\beta t_yL_x^{1-1/2K_x}.
\label{eq:rhoscale}
\end{equation}
The quantum phase transition occurring at a critical tunneling in 
the underlying Hubbard model corresponds to the well-known 
finite-temperature KT transition in the 2D XY-model. 
According to Weber-Minnhagen scaling \cite{Weber}, 
$\rho_{s,XY}$ should be 
proportional to $1+1/(2\ln L_y + C)$
at the critical point, where $C$ is a fitting constant to be
determined. The quantity $\rho_{s,XY}$ divided by this function should
therefore assume the same value for all system sizes at the critical
point, if $C$ and $K_x$ are chosen correctly. The best scaling is
obtained with $C=-1$ and $K_x=2$ and is shown in Fig.~\ref{fig:scale}
a). The scaling behavior supports the conclusion that the
transition is a KT transition and the critical point is $\tc/U\approx
0.27L_x^{-1.75}$.

An independent method for calculating the location of the transition 
is provided by fitting the correlation function in the
$y$-direction, $\Gamma_y$, to an algebraic decay function
according to \refeq{correlation}.  
The phase transition for a 1D 
Hubbard model at a fixed number of atoms per site is found by
determining the point where the exponent in the power law, $1/(2K_y)$,
becomes equal to 0.25 \cite{kuhner2000}.  


In \reffig{fig:scale} b) the exponent obtained from an algebraic fit
is displayed for different system sizes. There is a clear finite-size
effect, the larger $L_y$ the larger the critical value.  The critical
point at which $1/(2K_y)=0.25$ for the infinite system is determined
to be $\tc/U\approx 0.3L_x^{-1.75}$, in agreement with the scaling in
\reffig{fig:scale} a) and the point of loss of superfluidity observed
in \reffig{fig:Three}.

{\it Summary} -- We have reported theoretical evidence for a transition in a
2D Hubbard model between a 2D superfluid and
a 1D Mott insulating state consisting of isolated
1D tubes. The location of the transition is consistent with 
predictions based upon a random phase approximation made 
in Refs.\ \cite{ho2004,cazalilla2006,gangardt2006}.
By scaling it is shown that the system
undergoes a KT transition. 
The transition can be induced using atoms in an optical
lattice and observed by time-of-flight imaging.

This work was supported by the G\"oran Gustafsson foundation, the
Swedish Research Council, the Knut and Alice Wallenberg Foundation,  
the Carl Trygger foundation, and the Kempe foundation.



\end{document}